\documentclass[]{aa}  

\usepackage{graphicx}
\usepackage{txfonts}
\usepackage{natbib}

\begin{document}

\title{The effects of  metallicity on the Galactic  disk population of
  white dwarfs}

\author{Ruxandra Cojocaru\inst{1,2},
        Santiago Torres\inst{1,2}, 
        Jordi Isern\inst{3,2} \and
        Enrique Garc\'{\i}a--Berro\inst{1,2}}

\institute{Departament de F\'\i sica Aplicada,
           Universitat Polit\`ecnica de Catalunya,
           c/Esteve Terrades 5, 
           08860 Castelldefels, Spain\
           \and       
           Institute for Space  Studies of Catalonia,
           c/Gran Capit\`a 2--4, Edif. Nexus 104,   
           08034  Barcelona,  Spain\ 
           \and
           Institut de Ci\`encies de l'Espai (CSIC),
           Campus UAB, Facultat de Ci\`encies, Torre C-5
           08193 Bellaterra, Spain}

\offprints{E. Garc\'\i a--Berro}

\date{\today}

\abstract{It has been  known for a long time  that stellar metallicity
  plays a  significant role in  the determination  of the ages  of the
  different   Galactic  stellar   populations,   when  main   sequence
  evolutionary tracks are employed.}
{Here we  analyze the role that  metallicity plays on the  white dwarf
  luminosity function  of the  Galactic disk, which  is often  used to
  determine its age.}
{We employ a  Monte Carlo population synthesis code  that accounts for
  the properties of the population of Galactic disk white dwarfs.  Our
  code incorporates the most up-to-date evolutionary cooling sequences
  for   white  dwarfs   with   hydrogen-rich  and   hydrogen-deficient
  atmospheres for  both carbon-oxygen  and oxygen-neon cores.   We use
  two different models to assess the evolution of the metallicity, one
  in which the  adopted metallicity is constant with time,  but with a
  moderate  dispersion, and  a  second one  in  which the  metallicity
  increases with time.}
{We  found that  our theoretical  results are  in a  very satisfactory
  agreement with the observational  luminosity functions obtained from
  the Sloan  Digital Sky  Survey (SDSS) and  from the  SuperCOSMOS Sky
  Survey (SSS), independently of  the adopted age-metallicity law.  In
  particular, we found that the  age-metallicity law has no noticeable
  impact in  shaping the bright  branch of the white  dwarf luminosity
  function, and that the position of its cut-off is almost insensitive
  to the adopoted age-metallicity relationship.}
{Because the shape of the bright  branch of the white dwarf luminosity
  function is insensitive to the age-metallicity law, it can be safely
  employed to  test the theoretical evolutionary  sequences, while due
  to the  limited sensitivity of the  position of the drop-off  to the
  distribution  of  metallicities,  its  location  provides  a  robust
  indicator of the age of the Galactic disk.}


\titlerunning{The effects of the metallicity on the population of disk
  white dwafs}

\authorrunning{R. Cojocaru et al.}

\maketitle


\section{Introduction}

The evolutionary properties  of white dwarfs have been  the subject of
intensive studies over  the past three decades. This  is partially due
to the fact that these fossil stars convey important information about
the ages, formation and evolution of  a wide range of Galactic stellar
populations. In fact,  white dwarfs are the  evolutionary end-point of
most stars in  our Galaxy.  Specifically, white dwarfs  are the direct
descendants  of main-sequence  stars  with masses  ranging from  $\sim
0.8\, M_{\sun}$ to $\sim  10\, M_{\sun}$.  Actually, a straightforward
calculation shows that, given the  shape of the initial mass function,
over  98\% of  the stars  in  our Galaxy  that evolve  on a  timescale
younger than  its age  ($\sim 10$~Gyr) will  become white  dwarfs.  In
addition to  being so  numerous, white  dwarfs have  other interesting
properties.   Specifically,  the two  most  important  ones are  their
structural simplicity and their well known evolutionary properties ---
see,  e.g., \cite{Alt2010a}.  This, together  with the  fact that  the
evolution  of  white dwarfs  is  a  simple  and well  studied  cooling
process, in  which gravothermal  energy is  slowly radiated  away, has
allowed their use as independent  cosmic clocks.  In particular, white
dwarfs  have been  employed  to  estimate the  ages  of  a variety  of
Galactic   stellar   populations,   such    as   the   Galactic   disk
\citep{Win1987,Gar1988}  and halo  \citep{1990A&A...233..456M,Ise1998}
or  the system  of Galactic  globular \citep{Kal2001,Han2002,HansenGC}
and open clusters -- see, e.g., \cite{Gar2010} and \cite{Jeffery}, for
recent  studies.    Furthermore,  the   ensemble  properties   of  the
population of  Galactic disk white dwarfs  has also been used  to test
theoretical  scenarios  and theories  that  cannot  be probed  yet  in
terrestrial  laboratories.   These   include,  for  instance,  testing
alternative  theories of  gravitation which  result in  a hypothetical
variation of  the gravitational  constant \citep{gdot,  JCAP}, setting
constraints on the  mass of weakly interacting  particles, like axions
\citep{Isern}, or  constraining the  properties of the  so-called dark
forces \citep{2013PhRvD..88d3517D}.   Additionally, the  population of
white dwarfs has  also been used to derive  interesting constraints on
the local star formation  rate \citep{NohScalo1990, DPetal94, Rowell}.
Finally,  there is  a  plethora of  studies  dedicated to  extensively
analyze the contribution  of white dwarfs to the  baryonic dark matter
content  of   our  Galaxy   --  see,  for   instance,  \cite{Kawaler},
\cite{Tor2002}, \cite{Pauli} or \cite{Gar2004}, and references therein
-- a topic of the maximum interest.

In order to  obtain useful information from  the different populations
of white dwarfs, three conditions  must be fulfilled.  First, accurate
observational  data  are  needed.   With  the  advent  of  large-scale
automatic surveys,  like the Sloan Digital  Sky survey \citep{Gun1998}
or  the SuperCOSMOS  Sky Survey  \citep{Row2011} --  to cite  just two
representative examples  -- the sample  of white dwarfs  with reliable
and accurate measurements of their astronomical properties has largely
increased,  thus allowing  for detailed  comparisons with  theoretical
models.  The second important  ingredient to obtain useful information
from the  observed populations of  Galactic white  dwarfs is a  set of
accurate evolutionary  cooling sequences.  The degree  of accuracy and
sophistication  of  the  available  cooling sequences  has  also  been
considerably improved during the  last decade, and modern evolutionary
sequences   include   a   detailed  treatment   of   the   atmospheres
\citep{Rene}, as  well as an accurate  treatment of all the  sinks and
sources of energy in the deep interior of white dwarfs \citep{energy1,
  energy2}.  Among  other interesting  physical phenomena,  we mention
neutrino   emission,  $^{22}$Ne   diffusion   in   the  liquid   phase
\citep{neon1, neon2}, and phase separation of the carbon-oxygen binary
mixture upon crystallization \citep{segretain}.  Moreover, it is worth
emphasizing  that the  agreement between  the cooling  tracks computed
using different stellar evolutionary  codes is nowadays excellent, and
even   better  than   that  found   between  different   main-sequence
evolutionary  sequences   --  see  \cite{Salaris13}  for   a  detailed
discussion  of this  issue.  Finally,  a  tool to  model the  ensemble
properties of  these populations is  also needed.  For this,  the best
choice is  a Monte  Carlo simulator, a  technique which  is frequently
used to  model the different  Galactic stellar populations.   In fact,
Monte  Carlo  techniques  have  been  frequently  used  to  model  the
population of  disk white dwarfs,  compensating at times for  the poor
statistics of  the old observational samples  \citep{Gar1999, Gar2004}
and allowing  one to  study the  effects of biases  and of  the sample
selection procedure \citep{Geijo1, Geijo2}.

Cosmic age determinations are usually obtained by employing models and
evolutionary tracks  for stars  with Solar  metallicity, which  is the
median value  for the metallicity  distribution function (MDF)  of the
Galactic disk.  However the entire  span of this distribution function
is many times  neglected.  Over the years photometric  studies of main
sequence  stars  have resulted  in  the  release of  large  catalogues
\citep{Twarog1980,      1983A&AS...54...55O,      1987gal..proc..229S,
2004A&A...418..989N,  2008MNRAS.388.1175H}.  In  most occasions  these
catalogues  contain one  parameter  for metallicity,  and no  detailed
element  abundances. Perhaps,  one of  the most  important results  of
these surveys  is that  high-metallicity stars are  present throughout
the  entire mass  range,  and  that the  MDF  depends  on the  stellar
age. Even if the peak  value for the metallicity remains approximately
at Solar  values, young stars  have a much narrower  distribution than
old stars, which suggests that a correct sampling should be denser for
younger stars \citep{Cas2011}.  The rather  symmetric shape of the MDF
is   explained  through   the  natural   process  of   star  migration
\citep{Sel2002},  in which  high-metallicity  stars  migrate from  the
inner  disk and  low-metallicity  stars  do so  from  the outer  disk.
Additionally,  it is  believed that  old  stars probably  play a  more
significant role in this situation, given that they clearly contribute
to the  lower metallicity wing  of the MDF \citep{Cas2011}.   This may
have important consequences. On the one hand recent studies have shown
that high  values of the  metallicity imply slower  evolutionary rates
for moderately cool white dwarfs due to the sedimentation of $^{22}$Ne
\citep{neon1,  Gar2010,  neon2}.   On   the  other  hand,  metallicity
modifies the lifetime of white dwarf progenitors, and there is not yet
a consensus of whether the initial final mass relationship is modified
-- see  \cite{2005ASPC..334...43I}   for  a   preliminary  discussion.
Obviously,  these effects  may  affect the  age  determination of  the
Galactic disk  obtained using  the faint downturn  of the  white dwarf
luminosity function.  The ultimate goal of  this work is to assess how
the  evolution of  the metallicity  of the  Galactic disk  affects the
shape of the white dwarf luminosity function.

To accomplish this  aim we employ two metallicity laws.   The first of
them is  a metallicity law  with a  median corresponding to  the Solar
value, and a dispersion around  it \citep{Cas2011}, whereas the second
one    is    the    classical    age-metallicity    relationship    of
\cite{Twarog1980}.    The  paper   is   organized   as  follows.    In
Sect.~\ref{Code} we explain the most  basic features of our population
synthesis code,  while in  Sect.~\ref{data} we  describe the  two most
recent and  reliable observational samples  with which we  compare our
simulated data, and we discuss the selection criteria employed to cull
a  representative   sample  of   white  dwarfs  from   these  surveys.
Sect.~\ref{results} is devoted to explain  in depth the results of our
Monte  Carlo simulations,  and  to thoroughly  compare our  population
synthesis   results  with   the  observational   data.   Finally,   in
Sect.~\ref{conclusions},  we review  our  most  relevant findings,  we
discuss their significance, and we summarize our conclusions.


\section{The population synthesis code}
\label{Code}

A full  description of  all the relevant  ingredients employed  in our
Monte   Carlo  simulator   can  be   found  in   our   previous  works
\citep{Gar1999,Tor2002,Gar2004}. However, for the sake of completeness
we summarize here  its main inputs. The simulations  described in this
paper   were  done   using   a  random   number  generator   algorithm
\citep{James_1990} which provides a uniform probability density within
the interval $(0,1)$ and ensures a repetition period of $\ga 10^{18}$,
which is  virtually infinite for practical  simulations. When Gaussian
probability functions were needed  we used the Box-Muller algorithm as
described  in  \cite{NRs}.  

First, the  positions of  the stars were  randomly generated  within a
spherical region centered on the Sun  with a radius of $4.0$~kpc.  For
the local density of stars  a double exponential distribution is used,
with a constant  Galactic scale height of 250~pc and  a constant scale
length of 3.5~kpc.  The time at which each synthetic star was born was
generated according to a constant  star formation rate, and adopting a
galactic disk age of 9.5~Gyr, except as otherwise stated.  The mass of
each   star   follows   the   standard  initial   mass   function   of
\cite{Kro2001}.  Velocities were randomly obtained taking into account
the differential rotation of the  Galaxy, the peculiar velocity of the
Sun and  a dispersion  law which  depends on the  scale height  -- see
\cite{Gar1999} for details.  The  evolutionary ages of the progenitors
were interpolated in the BaSTI database of the appropriate metallicity
\citep{BaSTI}.  Given the  age of the Galaxy and  the age, metallicity
and mass of the progenitor star we know which synthetic stars have had
time to become white dwarfs, and  for these we derive their mass using
the  initial-final  mass  relationship of  \cite{Cat2008},  except  as
otherwise stated.

The cooling sequences employed in this  work depend on the mass of the
white  dwarf.  If  the mass  of the  white dwarf  is less  than $1.1\,
M_{\sun}$  a carbon-oxygen  core was  adopted, while  if the  mass was
greater than this value we assumed  that the core of the corresponding
white  dwarf  was   made  of  oxygen  and  neon.    In  our  reference
calculations we used the  evolutionary calculations of \citet{Ren2010}
for  carbon-oxygen white  dwarfs with  hydrogen-rich atmospheres,  and
those of \cite{DBs} for  hydrogen-deficient envelopes, while for white
dwarfs  with  oxygen-neon  cores  we used  those  of  \citet{Alt2007}.
However, to assess the effects of  the different cooling tracks in the
white dwarf luminosity function in  additional sets of calculations we
also employed  the cooling tracks  of \cite{Fon2001} and of  the BaSTI
project  \citep{Sal2010}  for  carbon-oxygen white  dwarfs  with  pure
hydrogen  atmospheres   and  those  of  \cite{Fon2001}   --  see  also
\cite{Ber2011} --  for white dwarfs with  helium dominated atmospheres
-- see below.  Using the  appropriate white dwarf evolutionary tracks,
we interpolated the luminosity, effective temperature and the value of
$\log g$ of  each synthetic star.  Additionally,  we also interpolated
their $UBVRI$  colors, which  we then converted  to the  $ugriz$ color
system, which is more adequate to compare our results with a sample of
white  dwarfs culled  from  the  SDSS (see  next  section), using  the
transformations described in \cite{Jordi}.  To compare with the sample
of white dwarfs of the SSS we used the color transformations described
in \citet{xu2007}.

For each  of the  models studied below  we generated  $50$ independent
Monte Carlo simulations (with different initial seeds) and for each of
these Monte Carlo  realizations, we increased the  number of simulated
Monte Carlo realizations to  $10^4$ using bootstrap techniques.  Using
this procedure  we ensure convergence in  all the final values  of the
relevant quantities. In what follows  for each quantity of interest we
present   the  ensemble   average   of  the   different  Monte   Carlo
realizations,  as  well  as   the  corresponding  standard  deviation.
Following  all  these  steps  we  were able  to  produce  a  synthetic
population of disk white dwarfs which, by definition, is complete.  To
this  population a  series of  filters,  which take  into account  the
observational cuts,  must be  applied. Describing  them in  detail is,
precisely, the goal of our next section.


\section{The observational samples}
\label{data}

The  two most  recent, complete  and  reliable samples  of disk  white
dwarfs are  those obtained from  the SDSS and  from the SSS.   In this
section we describe  them separately, placing special  emphasis on the
observational  cuts employed  to derive  the corresponding  luminosity
functions. In  our population  synthesis study  we follow  closely the
prescriptions  employed  to  obtain  the observed  samples,  with  the
ultimate  goal of  producing  theoretical white  dwarf populations  as
realistic as possible, for both the white dwarf sample of the SDSS and
for  that of  the SSS,  since this  is crucial  to derive  white dwarf
luminosity functions that can be compared to the observational data in
a meaningful way.

The  SDSS   surveyed  5,282~deg$^2$  of  high-latitude   sky  in  five
bandpasses $(ugriz)$ --  see \cite{Fukugita1996}, \cite{Gunn1998}, and
\cite{Gunn2006}  for additional  details  -- and  besides quasars  and
galaxies, obtained many spectra of  white dwarfs and other blue stars.
Using the SDSS  Data Release 4 \citet{Eis2006} presented  a catalog of
9,316 spectroscopically confirmed white  dwarfs.  The catalog contains
both hydrogen-rich (DA) and hydrogen-defficient (non-DA) white dwarfs,
and from  it \cite{Har2006}, using photometric  distances, USNO proper
motions, and the $1/V_{\max}$  method \citep{Sch1968}, derived a white
dwarf luminosity  function.  In particular, to  obtain this luminosity
function  they considered  all stars  brighter than  $g<19.5$, and  to
discrimate between main sequence stars  and white dwarfs they employed
the reduced proper motion
\begin{equation}
H_g=g+\log \mu+5=M_g+5\log V_{\rm tan}-3.379
\end{equation}
where $\mu$ is the proper  motion, and $V_{\rm tan}$ is the tangential
velocity. Specifically, they required that all objects contributing to
the white dwarf  luminosity function should be below  and blueward the
reduced proper motion corresponding to $V_{\rm tan}=20$~km~s$^{-1}$ in
the  reduced  proper motion-color  diagram.   Additionally, they  only
selected  white dwarfs  with $15<g<19.5$,  and proper  motions  $\mu >
0.02^{\prime\prime}$~yr$^{-1}$.  The resulting  sample  contains about
7,000 white dwarfs. Later  on, \cite{Deg2008} using 3,358 white dwarfs
presented an  improved white dwarf luminosity function,  in which only
spectroscopically confirmed DA white  dwarfs were employed.  To obtain
this  luminosity function  they adopted  $V_{\rm tan}>30$~km~s$^{-1}$.
However,  given  that this  luminosity  function  does  not present  a
cut-off, as the luminosity function  of \cite{Har2006} does, we do not
use it.

The  other large  observational sample  of disk  white dwarfs  is that
obtained  from  the  SSS  \citep{Row2011}.   The SSS  is  an  advanced
photographic plate digitizing machine.  The plates were taken with the
UK  Schmidt  telescope (UKST),  the  ESO  Schmidt  telescope, and  the
Palomar  Schmidt  telescope,  and  the resulting  catalogs  have  been
compiled  by digitizing  several generations  of  photographic Schmidt
plates.  The photometric system of this survey is less well known than
that of the  SDSS and has three passbands  $(b_{\rm J}, r_{59{\rm F}},
i_{\rm N})$ -- see \cite{SSS}  for details.  Using the data of several
generations of  these plates  \cite{Row2011} constructed a  catalog of
about $10,000$ white dwarfs with magnitudes down to $r_{59{\rm F}}\sim
19.75$     and     proper    motions     as     low    as     $\mu\sim
0.05^{\prime\prime}$~yr$^{-1}$, which covers  nearly three quarters of
the sky.   The observational selection criteria adopted  to derive the
corresponding  white  dwarf  luminosity  function are  the  following.
First, the proper motion cut depends on the $b_{\rm J}$ magnitude
\begin{equation}
\mu > 5\left(\sigma_\mu^{\max}(b_{\rm J})  +  0.002\right)
\end{equation} 
In  this expression  $\sigma_\mu$  is the  standard  deviation in  the
proper  motion  measurements, which  means  that  the measured  proper
motion  is $5\sigma$ larger  than the  error at  a given  $b_{\rm J}$.
Additionally, there is  a magnitude cut $12 <  r_{59{\rm F}} < 19.75$,
whereas they also imposed a  cut in the reduced proper motion diagram,
adopting $V_{\rm tan}>30$~km~s$^{-1}$.


\section{Results}
\label{results}

Before  assessing  the role  of  the  adopted  metallicity law  it  is
convenient to check  whether or not other aspects can  mask its impact
on the  white dwarf luminosity  function.  For instance, the  ratio of
white  dwarfs  with  hydrogen-deficient   atmospheres  to  those  with
hydrogen-rich  ones depends  on  the effective  temperature, and  this
could  possibly influence  the  shape of  the  white dwarf  luminosity
function. Hence, we first check if  this could mask the effects of the
adopted metallicity law.  This is  done in Sect.~\ref{DB/DA}. The same
can  be  said  about  the  adopted  theoretical  white  dwarf  cooling
sequences. Quite obviously the adopted white dwarf cooling tracks also
influence the precise shape of the luminosity function.  Consequently,
in   Sect.~\ref{tracks}   we   also    explore   and   quantify   this
possibility. Finally, in  Sect.~\ref{Z} we analyze the  effects of the
adopted  metallicity  law.   We  emphasize  that  all  the  luminosity
functions  presented in  this section  have been  computed adopting  a
constant star formation rate.

\subsection{The fraction of non-DA white dwarfs}
\label{DB/DA}

To begin with, we discuss the role of the ratio of the number of white
dwarfs  with hydrogen-deficient  atmospheres  to the  total number  of
white dwarfs,  including those  with hydrogen-rich  atmospheres.  Most
white  dwarf stars  have hydrogen-dominated  atmospheres, constituting
the group  of so-called DA white  dwarfs.  However, hydrogen-deficient
white  dwarfs, known  as non-DA  stars, represent  between $15\%$  and
$25\%$   of  the   total  white   dwarf  population   of  the   Galaxy
\citep{Fon2001}. In our attempt to properly analyze factors that might
potentially  influence  the  shape   of  the  white  dwarf  luminosity
function,  and  given  that   the  most  recent  luminosity  functions
incorporate both DA and non-DA white dwarfs, we first try to reproduce
the observed ratio of non-DA to the total number of white dwarfs.  For
this  set  of  simulations  we  adopt the  cooling  sequences  of  our
reference model and  the evolutionary ages of  white dwarf progenitors
of Solar  metallicity. That  is, we assumed  that all  synthetic stars
have Solar metallicity.

\begin{figure}
   \resizebox{\hsize}{!}
   {\includegraphics[width=\columnwidth]{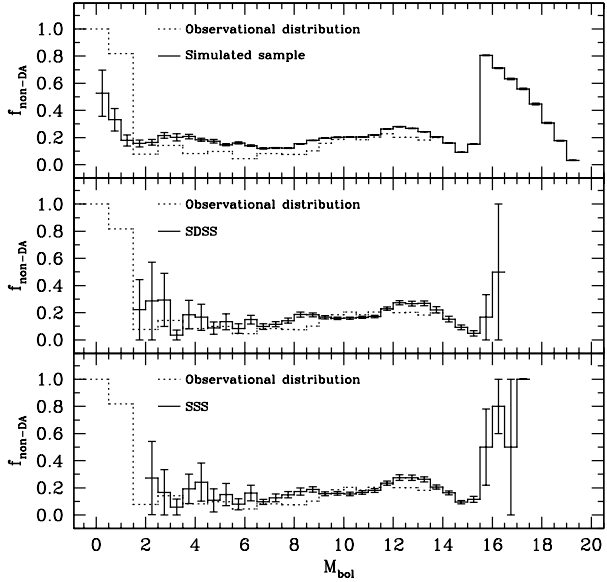}}
   \caption{Distribution of the ratio of non-DA to the total number of
     white  dwarfs as  a function  of the  bolometric magnitude.   The
     solid line  represents the  simulated sample, whereas  the dotted
     line  corresponds  to  the   observational  data.  See  text  for
     additional details.}
\label{fig:dist}
\end{figure}

The  observed ratio  of  non-DA  white dwarfs  as  a  function of  the
bolometric   magnitude    is   shown   in   all    three   panels   of
Fig.~\ref{fig:dist} using  a dotted line.  The  observational data has
been  compiled from  \cite{Krz2009} for  hot white  dwarfs --  namely,
those  with  effective  temperatures larger  than  $T_{\rm  eff}\simeq
25,000$~K  -- and  \cite{Deg2008}  and references  therein for  cooler
white dwarfs.  The upper panel  of Fig.~\ref{fig:dist} shows the ratio
of the number  of non-DA white dwarfs as a  function of the bolometric
magnitude  when all  synthetic white  dwarfs of  an otherwise  typical
Monte  Carlo realization  are considered  -- that  is, for  a complete
sample -- as a solid line.  This simulation was computed assuming that
at birth  20\% of  white dwarfs have  atmospheres devoid  of hydrogen,
whilst the lower two panels  show, respectively, the same distribution
when the cuts of \cite{Har2006}  and \cite{Row2011} are applied to the
complete  sample.   Thus,  these  two last  Monte  Carlo  realizations
incorporate the biases introduced by the selection criteria. As can be
seen, in  general our  theoretical resuls  are in  excellent agreement
with  the observational  data,  except at  very  high luminosities  --
namely  for  $M_{\rm  bol}\la  2.0$  -- for  which  the  non-DA  ratio
predicted by our  simulations for the complete  sample is considerably
smaller ($\sim  0.5$) than  the observed one  ($\sim 1.0$).   We note,
however, that  at these  bolometric magnitudes  the number  density of
white  dwarfs is  very small  and,  hence, this  ratio presents  large
fluctuations,  depending  on  the  specific  Monte  Carlo  realization
chosen.  It is worth noting as  well that the effects of the selection
criteria for this range of  bolometric magnitudes is important, as for
the complete sample  the ratio of non-DA white dwarfs  turns out to be
$\sim  0.5$, whereas  this  ratio  drops to  $\sim  0.0$  for the  two
restricted  samples, although  the statistical  fluctuations are  very
large. Another interesting  aspect of the theoretical  results that is
also worth commenting is that  at very low luminosities our population
synthesis calculations predict  a noticeable increase of  the ratio of
non-DA to DA  white dwarfs for the complete sample,  in agreement with
what is found for the local sample of white dwarfs -- see \cite{local}
and references  therein.  This is  evident as  well for the  sample of
white dwarfs  in which  the selection  criteria of  \cite{Row2011} are
used.  If the  selection criteria of \cite{Har2006}  are employed, the
ratio of non-DA  white dwarfs does not show such  a markedly increase,
because very few  low luminosity white dwarfs are found  when this set
of  criteria  is adopted.   We  will  elaborate  on this  below,  when
discussing the white dwarf luminosity function.

\begin{figure}
   \resizebox{\hsize}{!}
   {\includegraphics[width=\columnwidth]{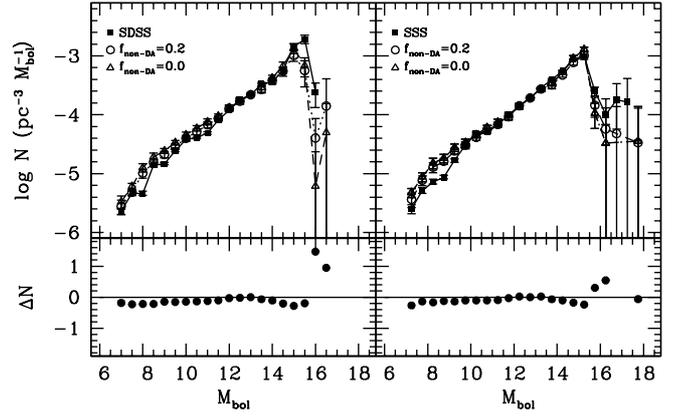}}
   \caption{Simulated white  dwarf luminosity  function for a  pure DA
     white dwarf population  (open triangles) and for  a population in
     which a canonical  fraction of 20\% of non-DA  white dwarfs (open
     circles) is adopted. Both  luminosity functions are compared with
     the observational white dwarf luminosity function (solid squares)
     of \cite{Har2006} (left top  panel) and \cite{Row2011} (right top
     panel).  In the  bottom panels we show the  residuals between the
     synthetic luminosity functions.}
\label{fig:fldb}
\end{figure}

Once established that the adopted  $20\%$ ratio of non-DA white dwarfs
at birth results  in a credible distribution for  all luminosities, we
test the  actual impact of  this ratio  on the white  dwarf luminosity
function. To  this end  we compute  an additional  set of  Monte Carlo
simulations for which we adopt $f_{\rm non-DA}=0.0$ and we compare the
corresponding  luminosity function  with that  obtained using  $f_{\rm
non-DA}=0.2$.    The   result   of   this   exercise   is   shown   in
Fig.~\ref{fig:fldb}.  In the top panels  of this figure the luminosity
functions  obtained when  $f_{\rm non-DA}=0.0$  -- open  triangles and
dotted-dashed  line  -- and  $f_{\rm  non-DA}=0.2$  -- hollow  circles
connected  by a  dashed line  -- are  compared, while  the solid  line
connecting the solid squares  display the observed luminosity function
of \cite{Har2006} -- left panel  -- and \cite{Row2011} -- right panel.
Obviously,  to produce  these luminosity  functions the  corresponding
selection criteria have  been used.  The bottom panels  of this figure
display the  residuals between both theoretical  luminosity functions,
which  better   help  in   assessing  the  differences   between  both
simulations, defined as:

\begin{equation}
\Delta N = 2\frac{N_{f_{\rm DB}=0.2}-N_{f_{\rm DB}=0}}
                 {N_{f_{\rm DB}=0.2}+N_{f_{\rm DB}=0}}
\end{equation}

As can be seen in the bottom panels of figure Fig.~\ref{fig:fldb}, the
effect  of the  adopted  non-DA  ratio on  the  simulated white  dwarf
luminosity  functions is  minimal,  except  for bolometric  magnitudes
larger  than $M_{\rm  bol}\simeq  15$. For  the  bins with  bolometric
magnitudes larger  than this value  the differences are  noticeable in
both   cases,  being   the  luminosity   function  in   which  $f_{\rm
non-DA}=0.2$  in  better agreement  with  the  observational one.   In
particular, when $f_{\rm non-DA}=0.2$ is adopted the number density of
low-luminosity white dwarfs increases, as should be expected, given
that white  dwarfs with  hydrogen-defficient atmospheres  cool faster.
Actually,  the  number density  of  white  dwarfs for  the  luminosity
functions computed with $f_{\rm  non-DA}=0.0$ is consistent with zero,
since for these bins in all Monte Carlo realizations we obtain at most
only one object per  bin. We thus conclude that the  hot branch of the
white dwarf luminosity  function is almost insensitive  to the adopted
ratio  of non-DA  white dwarfs,  and that  the effects  of this  ratio
concentrate   in  the   poorly   determined  bins   with  the   lowest
luminosities.

\begin{figure}
   \resizebox{\hsize}{!}
   {\includegraphics[width=0.9\columnwidth]{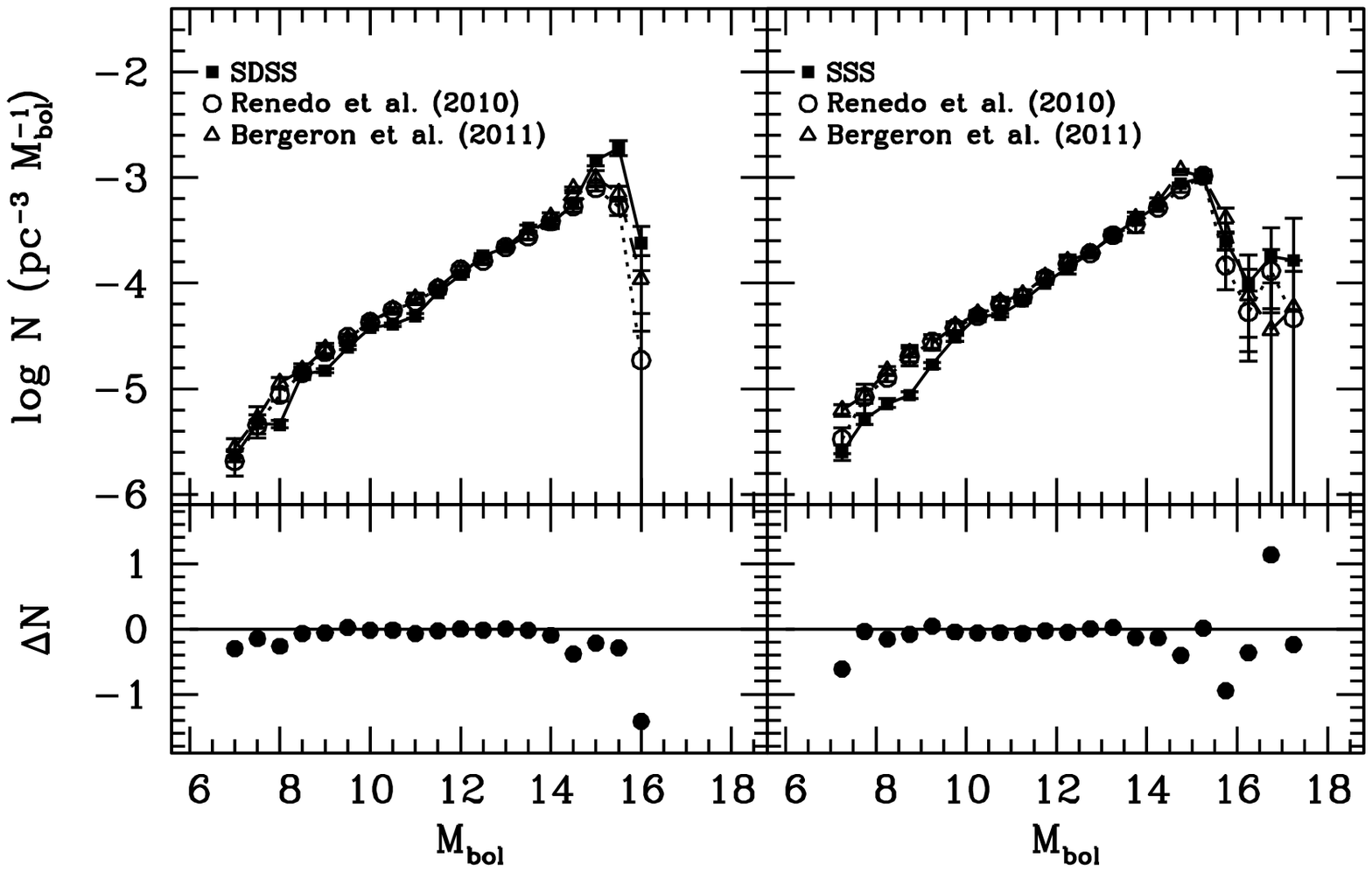}}\\
   \resizebox{\hsize}{!}
   {\includegraphics[width=0.9\columnwidth]{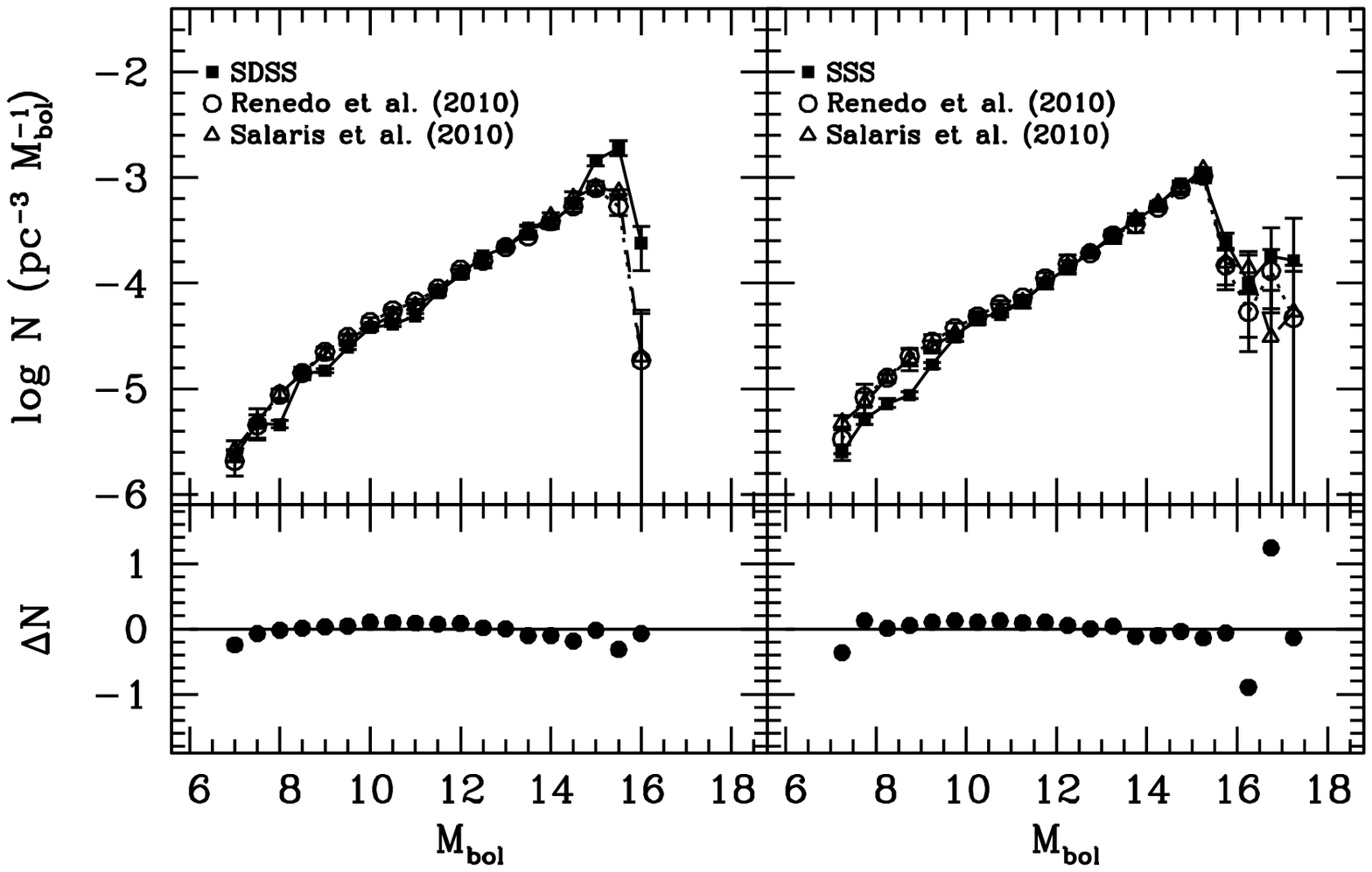}}
   \caption{Simulated white dwarf  luminosity functions computed using
     our reference model -- open circles -- compared to those obtained
     using the cooling sequences of \cite{Fon2001} -- top panels, open
     triangles  --  and \cite{Sal2010}  --  bottom  panels, also  open
     triangles.  Also   shown  are   the  observational   white  dwarf
     luminosity  function (solid  squares) of  \cite{Har2006} --  left
     panels  -- and  \cite{Row2011}  -- right  panels.  The  synthetic
     white  dwarf luminosity  functions  have been  normalized to  the
     observational point with minimum error bars. }
\label{fig:MC123}
\end{figure}

\subsection{The effects of the adopted cooling tracks}
\label{tracks}

We  now discuss  which set  of cooling  sequences best  reproduces the
shape of the observed white dwarf  luminosity function. To this end in
Fig.~\ref{fig:MC123} we  compare various sets of  evolutionary cooling
sequences with  the two most  recent and reliable  observational white
dwarf  luminosity functions,  namely  that of  \cite{Har2006} --  left
panels -- and that of \cite{Row2011}  -- right panels.  The data shown
in this figure  corresponds to the ensemble average  of several single
independent Monte  Carlo realizations,  as described before,  with the
inputs  described in  Sect.~\ref{Code}.   In the  top  panels of  this
figure we display  the results of such comparison  when the luminosity
functions  obtained using  the  cooling tracks  of \cite{Fon2001}  are
compared  to  those   obtained  using  our  reference   model  --  see
Sect.~\ref{Code} -- whereas in the  bottom panels the results obtained
when the  cooling sequences  of \cite{Sal2010}  are compared  to those
obtained when the  cooling tracks of \cite{Ren2010} are  used. In each
panel we also show the residuals between both theoretical simulations.
As we did previously, for all the calculations shown in this figure we
adopted a constant metallicity equal to the Solar value.

Fig.~\ref{fig:MC123}  demonstrates  that  all three  sets  of  cooling
sequences  match very  well the  bright branch  of both  observational
white dwarf luminosity  functions, and that there  are not appreciable
differences   between  all   three   luminosity   functions  at   high
luminosities.   This agreement  is even  better when  the white  dwarf
luminosity function of  \cite{Row2011} is adopted, as  the location of
the maximum of the white dwarf luminosity function is nicely fitted by
all three  sets of cooling  tracks. Now we  turn our attention  to the
cool branch of  the white dwarf luminosity function, and  the shape of
its  drop-off. Again,  all  three  sets of  cooling  sequences are  in
excellent agreement with the observational data, although we emphasize
that at these  very low luminosities selection effects may  play a key
role  and both  the Monte  Carlo simulated  sample and  the real  ones
suffer from  poor statistics.   Nevertheless, we  note that  all three
sets of calculations  reproduce the downturn at  very faint magnitudes
of the luminosity function obtained from the SSS, which is not present
in the  luminosity function of  \cite{Har2006}, because of  the sample
selection  procedures.  Given  these  results there  are no  objective
reasons for  discarding any set  of evolutionary sequences,  and hence
for  the rest  of the  paper  we adopt  the cooling  sequences of  our
fiducial model, which is based on the theoretical cooling sequences of
\cite{Ren2010}.

\subsection{The influence of the metallicity law}
\label{Z}

Metallicity  is  known  to  have   a  considerable  influence  in  age
estimations  obtained when  the  turn-off of  main  sequence stars  of
stellar clusters is employed to  derive the corresponding age.  Hence,
it  is natural  to wonder  whether  or not  the age  derived from  the
drop-off of  the disk white  dwarf luminosity function depends  on the
adopted  age-metallicity relationship.   To perform  this analysis  we
adopted two different metallicity laws. The first one is the classical
metallicity  law   of  \cite{Twarog1980},  which  predicts   that  the
metallicity of  white dwarf  progenitor stars  monotonically increases
from zero  at very early ages  of our Galaxy to  Solar metallicity for
the current age of the Galactic disk.  Specifically, the metallicities
of  the individual  synthetic  stars are  randomly  drawn following  a
Gaussian distribution centered in $F(t)$, where $F(t)$ is a polynomial
fit to  the results  presented in Fig.~1  of \cite{Bravo1993},  with a
dispersion $\sigma=0.1$.   Our second age-metallicity  relationship is
based  on the  observational data  compiled by  the Geneva-Conpenhague
survey  \citep{Cas2011}.    This  recent   study  predicts   that  the
metallicity  of the  Galactic disk  is constant  with time  but has  a
sizable dispersion.   We note that this  distribution of metallicities
has been confirmed by the  recent results of \cite{Dur2013}. Thus, for
the  same  age   white  dwarf  progenitors  span  a   broad  range  of
metallicities.  In particular, for each synthetic star we employ ${\rm
[Fe/H]=[Fe/H]}_{\sun}+\Delta{\rm [Fe/H]}$, where $\Delta {\rm [Fe/H]}$
is randomly  drawn from a  Gaussian distribution with  the metallicity
spread found by \cite{Cas2011}, $\sigma=0.4$.

Whatever the adopted age-metallicity relationship is, in both cases we
interpolate both the  properties of the progenitors --  that is, their
main-sequence lifetimes -- and those  of the resulting white dwarfs --
namely, their  final masses,  cooling times, colors  and luminosities.
For  this  analysis   we  adopt  the  set  of   cooling  sequences  of
\cite{Ren2010}, which  almost cover  the full range  of metallicities.
If a  progenitor star  has metallicity lower  than $Z=10^{-3}$  -- the
lowest metallicity of the  evolutionary sequences of \cite{Ren2010} --
we  adopt  the  values  predicted   by  these  calculations.   If  the
metallicity of  the progenitor  stars is  larger than  $Z=10^{-2}$, we
interpolate between the evolutionary  sequences of \cite{Ren2010}, and
those of \cite{neon2},  which have a maximum  metallicity of $Z=0.06$,
which is  enough for our  purposes. Finally, we  adopt an age  for the
Galactic disk of 9.5~Gyr.

\begin{figure}
   \resizebox{\hsize}{!}
   {\includegraphics[width=0.9\columnwidth]{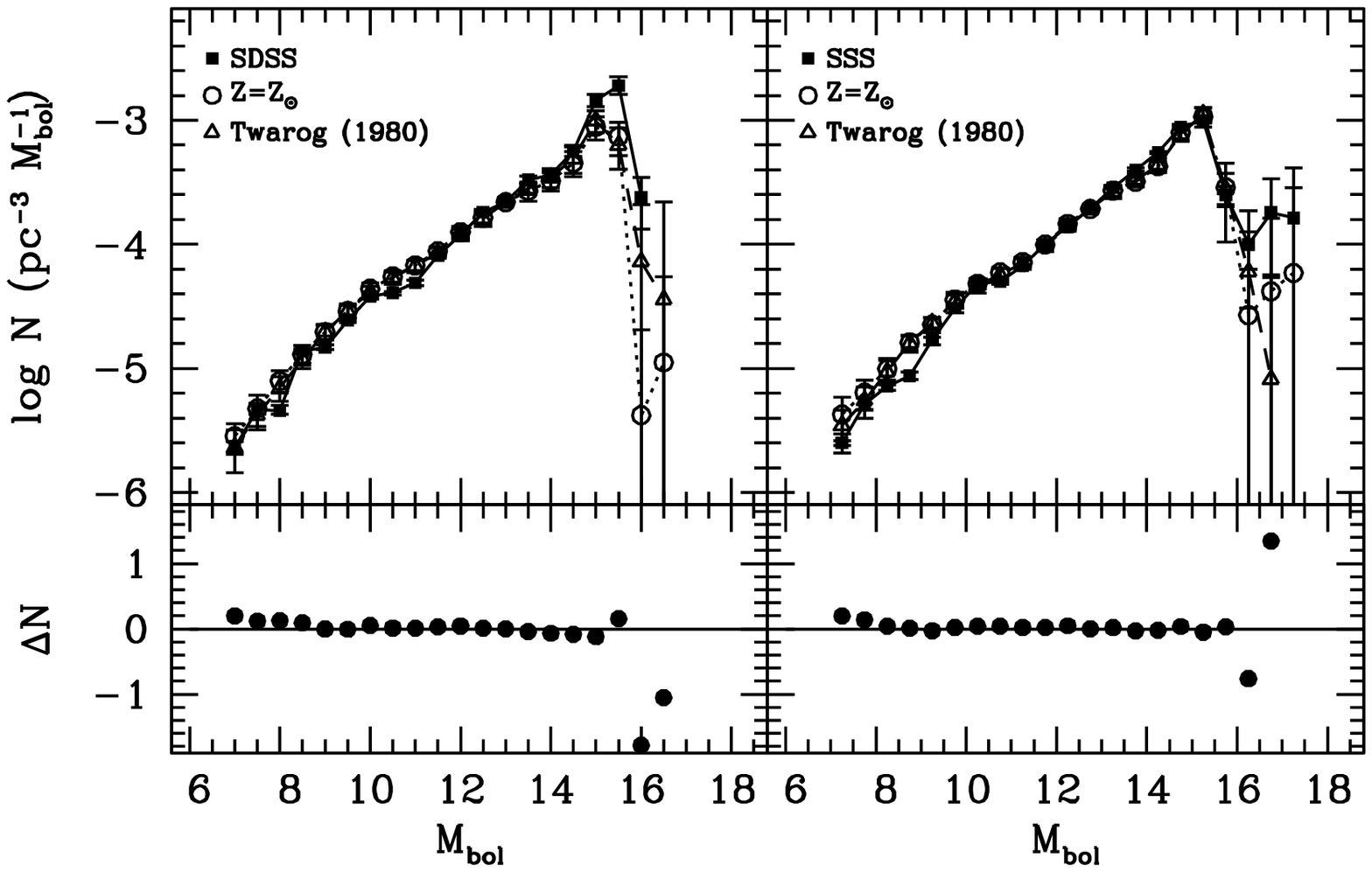}}\\
   \resizebox{\hsize}{!}
   {\includegraphics[width=0.9\columnwidth]{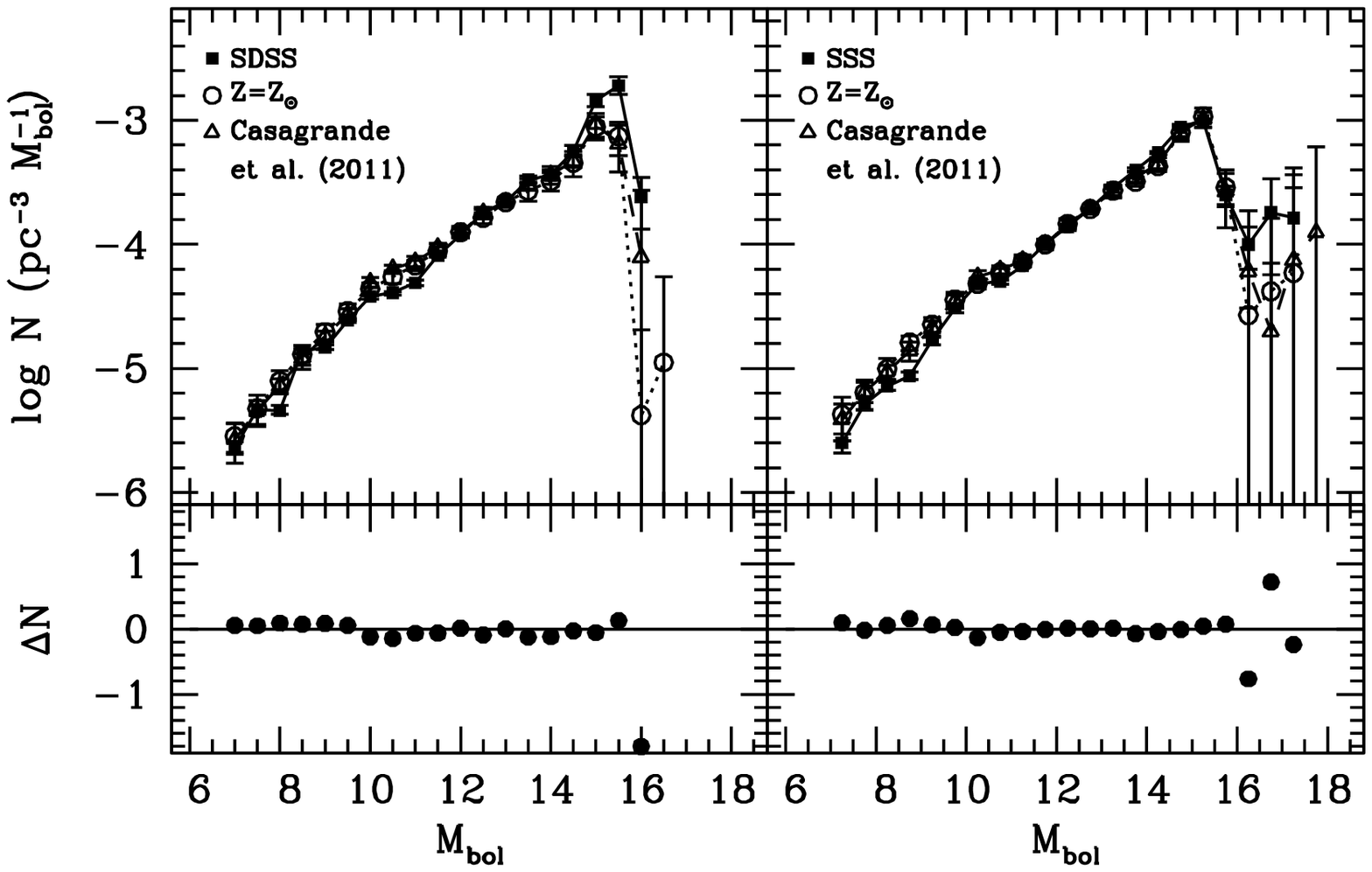}}
   \caption{Simulated  white  dwarf   luminosity  function  using  the
     metallicity law of \cite{Twarog1980} -- top panels -- and that of
     \cite{Cas2011} -- bottom panels  -- compared to the observational
     white dwarf luminosity functions of \cite{Har2006} -- left panels
     -- and \cite{Row2011} -- right panels.  As in previous figures we
     also  show   the  residuals  between  the   synthetic  luminosity
     functions  computed  with  a   given  metallicity  law,  and  the
     luminosity function  computed adopting the Solar  value (see text
     for details).}
\label{fig:flmet}
\end{figure}

Our results are presented  in Fig.~\ref{fig:flmet}.  In particular, in
the upper  panels of this  figure we compare the  luminosity functions
obtained using  the age-metallicity relationship  of \cite{Twarog1980}
-- open triangles -- and using a  fixed value of the metallicity equal
to the Solar one -- open  circles -- with the observational luminosity
functions of the SDSS -- left panels,  solid squares -- and the SSS --
right panels,  solid squares -- whereas  in the bottom panels  we show
the  same comparison,  but  this  time using  the  metallicity law  of
\cite{Cas2011}. As in previous figures  we also show the corresponding
differences between the  theoretical calculations.  As can  be seen in
this  figure,  both theoretical  calculations  yield  almost the  same
results, and most importantly, the position of the theoretical cut-off
of the white dwarf luminosity function  is not affected by the adopted
metallicity law, no  matter if the adopted metallicity law  is that of
\cite{Twarog1980}, that of \cite{Cas2011}, or  if a fixed value of the
metallicity is adopted. This, in turn,  means that the age estimate of
the Solar neighborhood  obtained from the location of  the drop-off of
the  white dwarf  luminosity function  is robust.   Actually, the  age
difference between the theoretical  calculations is rather small.  For
instance,  when the  luminosity function  of  the SDSS  is adopted  we
obtain that the  age of the disk derived using  the metallicity law of
\cite{Cas2011}  is   9.5~Gyr,  whereas  if  the   metallicity  law  of
\cite{Twarog1980} is  used the age  turns out to  be 10~Gyr, and  if a
fixed value  of the metallicity equal  to the Solar one  is adopted we
obtain 9.8~Gyr.   These ages when  the adopted luminosity  function is
that of the SSS  are 10~Gyr for the first two  cases, and 10.3~Gyr for
the case in  which a constant Solar metallicity with  no dispersion is
employed.

\begin{figure}
   \resizebox{\hsize}{!}
   {\includegraphics[width=0.9\columnwidth]{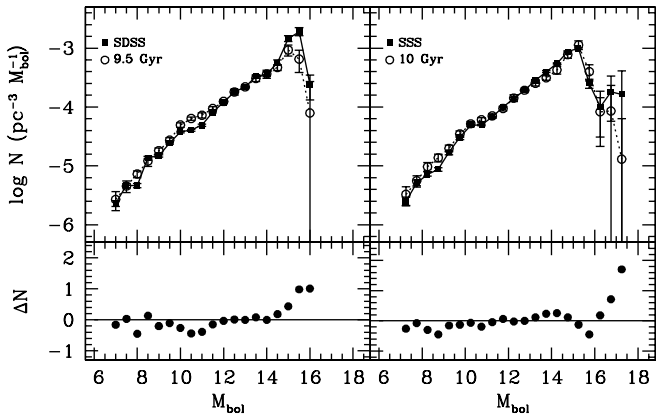}}
   \caption{Simulated white dwarf luminosity  functions using the very
     recent  metallicity  law  of   \cite{Cas2011},  compared  to  the
     observed luminosity  functions of the  SDSS (left panel)  and the
     SSS (right  panel). The adopted  disk ages are 9.5  and 10.0~Gyr,
     respectively, which are the ages that best reproduce the position
     of  the  observed  cut-off  of the  luminosity  function.  As  in
     previous figures we also show the corresponding residuals.}
\label{fig:flmodels}
\end{figure}

At first look this result may  seem surprising since at the luminosity
of  the cut-off  contribute the  oldest  white dwarfs  in the  Galaxy.
These white dwarfs have cooling  ages which are approximately equal to
the age of  the Galactic disk, and thus  have negligible main-sequence
lifetimes,   which    for   the   age-metallicity    relationship   of
\cite{Twarog1980} translates  in small metallicities.   However, since
these  white dwarfs  have progenitors  with very  short main  sequence
lifetimes   it  also   means   that  their   progenitors  are   rather
massive. Consequently,  the steep slope  of the initial  mass function
prevents  the formation  of a  large number  of these  stars.  On  the
contrary, the peak of the white dwarf luminosity function is dominated
by white dwarfs of $\sim 0.6\pm  0.05 \, M_{\sun}$, which are the bulk
of  white  dwarfs in  our  Solar  neighborhood,  and which  have  very
different main sequence lifetimes (and metallicities).  The net result
is that  the drop-off of the  white dwarf luminosity function  is very
abrupt, and  its position  is mainly  determined by  otherwise typical
white dwarfs.  Additionally, for  the case of  the metallicity  law of
\cite{Cas2011},  the  relatively  large  dispersion  of  metallicities
contributes to erase the  metallicity dependence of stellar lifetimes,
and results in a negligible dependence of the position of the drop-off
of the  white dwarf  luminosity function on  the metallicity  of white
dwarf progenitors.

Given that the position of the  cut-off of the theoretical white dwarf
luminosity function does not depend on the adopted metallicity law, we
computed the age of the disk  that best fits the observed distribution
of white  dwarfs. The resulting  white dwarf luminosity  functions are
displayed in  Fig.~\ref{fig:flmodels}, and the corresponding  ages are
9.5~Gyr for the  case in which the white dwarf  luminosity function of
the SDSS  is employed, whilst  we obtain 10.0~Gyr when  the luminosity
function of the SSS is adopted. Thus, currently, the age determination
of the Galactic  disk using the cut-off of the  white dwarf luminosity
function seems to be dominated by the relatively small number of white
dwarfs in  the lowest luminosity bins.   A comparison of the  SDSS and
SSS luminosity functions indicates that this introduces an uncertainty
of $\sim 0.5$~Gyr.


\section{Summary and conclusions}
\label{conclusions}

In this paper  we explored the possible dependence of  the position of
the  white dwarf  luminosity function  on the  adopted age-metallicity
relationship. Our  motivation for  undertaking such  a study  was that
most  age  estimators  depend  on   the  metallicity  of  the  adopted
theoretical stellar evolutionary models, and  such a study was lacking
for the case in which the age  of the Galactic disk is estimated using
the position  of the cut-off  of the white dwarf  luminosity function.
Our study  fills this gap.  In  doing so, we used  an up-to-date Monte
Carlo  simulator,  which incorporates  the  most  recent and  reliable
prescriptions for both  the main sequence lifetimes  and cooling ages,
as well  as the most  realistic Galactic  inputs. We also  adopted two
typical, and frequently used metallicity laws, the first one being the
classical  age-metallicity  relationship of  \cite{Twarog1980},  which
displays a monotonically  incresing trend, and the  metallicity law of
\cite{Cas2011}, which does not increase as time passes by, but instead
shows  a   relatively  large   dispersion  around  the   Solar  value.
Nevertheless, before studying the  role of the adopted age-metallicity
relationship,  we  also  studied   other  possible  effects  that  may
eventually have  a noticeable impact on  the shape of the  white dwarf
luminosity function.  Specifically, we studied how the ratio of non-DA
to DA white dwarfs influences the shape of the luminosity function and
of its  cut-off, and also  how the  choice of the  adopted theoretical
degenerate cooling sequences affects the luminosity function. We found
that  neither the  shape  of the  bright portion  of  the white  dwarf
luminosity  function  nor   the  position  of  its   downturn  at  low
luminosities   are  noticeable   affected  by   these  inputs.    More
importantly,   we  also   found  that   the  adopted   age-metallicity
relationship has  a negligible impact  on the shape of  the luminosity
function and on the position of its drop-off. Hence, the age estimates
of the Galactic disk obtained from the number counts of low-luminosity
white dwafs are robust, and the age discrepancies mainly come from the
way in which low-luminosity white dwarfs are culled to obtain a nearly
complete sample.  For the two most recent observational determinations
of   the  white   dwarf  luminosity   function  --   namely  that   of
\cite{Har2006}, which was obtained from white dwarfs found in the SDSS
-- and that of  \cite{Row2011}, which was derived using  data from the
SSS -- we obtain, respectively 9.5~Gyr and 10.0~Gyr.


\begin{acknowledgements}
This research  was supported by  AGAUR, by MCINN  grant AYA2011–23102,
and by the European Union  FEDER funds. RC also acknowledges financial
support from the FPI grant with reference BES-2012-053448.
\end{acknowledgements}


\bibliographystyle{aa}
\bibliography{Fe-H}

\begin{thebibliography}{66}
\expandafter\ifx\csname natexlab\endcsname\relax\def\natexlab#1{#1}\fi

\bibitem[{{Althaus} {et~al.}(2010{\natexlab{a}}){Althaus}, {C\'{o}rsico},
  {Isern}, \& {Garc\'{i}a-Berro}}]{Alt2010a}
{Althaus}, L.~G., {C\'{o}rsico}, A.~H., {Isern}, J., \& {Garc\'{i}a-Berro}, E.
  2010{\natexlab{a}}, \aapr, 18, 471

\bibitem[{Althaus {et~al.}(2007)Althaus, {Garc{\'{\i}}a-Berro}, {Isern},
  {C{\'o}rsico}, \& {Rohrmann}}]{Alt2007}
Althaus, L.~G., {Garc{\'{\i}}a-Berro}, E., {Isern}, J., {C{\'o}rsico}, A.~H.,
  \& {Rohrmann}, R.~D. 2007, \aap, 465, 249

\bibitem[{{Althaus} {et~al.}(2010{\natexlab{b}}){Althaus},
  {Garc{\'{\i}}a-Berro}, {Renedo}, {Isern}, {C{\'o}rsico}, \&
  {Rohrmann}}]{neon2}
{Althaus}, L.~G., {Garc{\'{\i}}a-Berro}, E., {Renedo}, I., {et~al.}
  2010{\natexlab{b}}, \apj, 719, 612

\bibitem[{{Benvenuto} \& {Althaus}(1997)}]{DBs}
{Benvenuto}, O.~G. \& {Althaus}, L.~G. 1997, \mnras, 288, 1004

\bibitem[{{Bergeron} {et~al.}(2011){Bergeron}, {Wesemael}, {Dufour},
  {Beauchamp}, {Hunter}, {Saffer}, {Gianninas}, {Ruiz}, {Limoges}, {Dufour},
  {Fontaine}, \& {Liebert}}]{Ber2011}
{Bergeron}, P., {Wesemael}, F., {Dufour}, P., {et~al.} 2011, \apj, 737, 28

\bibitem[{{Bravo} {et~al.}(1993){Bravo}, {Isern}, \& {Canal}}]{Bravo1993}
{Bravo}, E., {Isern}, J., \& {Canal}, R. 1993, \aap, 270, 288

\bibitem[{{Casagrande} {et~al.}(2011){Casagrande}, {Sch\"{o}nrich}, {Asplund},
  {Cassisi}, {Ram\'{i}rez}, {Mel\'{e}ndez}, {Bensby}, \& {Feltzing}}]{Cas2011}
{Casagrande}, L., {Sch\"{o}nrich}, R., {Asplund}, M., {et~al.} 2011, \aap, 530,
  A138

\bibitem[{{Catal\'{a}n} {et~al.}(2008){Catal\'{a}n}, {Isern},
  {Garc\'{i}a-Berro}, \& {Ribas}}]{Cat2008}
{Catal\'{a}n}, S., {Isern}, J., {Garc\'{i}a-Berro}, E., \& {Ribas}, I. 2008,
  \mnras, 387, 1693

\bibitem[{{De Gennaro} {et~al.}(2008){De Gennaro}, {von Hippel}, {Winget},
  {Kepler}, {Nitta}, {Koester}, \& {Althaus}}]{Deg2008}
{De Gennaro}, S., {von Hippel}, T., {Winget}, D.~E., {et~al.} 2008, \aj, 135, 1

\bibitem[{{Diaz-Pinto} {et~al.}(1994){Diaz-Pinto}, {Garcia-Berro}, {Hernanz},
  {Isern}, \& {Mochkovitch}}]{DPetal94}
{Diaz-Pinto}, A., {Garcia-Berro}, E., {Hernanz}, M., {Isern}, J., \&
  {Mochkovitch}, R. 1994, \aap, 282, 86

\bibitem[{{Dreiner} {et~al.}(2013){Dreiner}, {Fortin}, {Isern}, \&
  {Ubaldi}}]{2013PhRvD..88d3517D}
{Dreiner}, H.~K., {Fortin}, J.-F., {Isern}, J., \& {Ubaldi}, L. 2013, \prd, 88,
  043517

\bibitem[{{Duran} {et~al.}(2013){Duran}, {Ak}, {Bilir}, {Karaali}, {Ak},
  {Bostanc{\i}}, \& {Co{\c s}kuno{\u g}lu}}]{Dur2013}
{Duran}, {\c S}., {Ak}, S., {Bilir}, S., {et~al.} 2013, \pasa, 30, 43

\bibitem[{{Eisenstein} {et~al.}(2006){Eisenstein}, {Liebert}, {Harris},
  {Kleinman}, {Nitta}, {Silvestri}, {Anderson}, {Barentine}, {Brewington},
  {Brinkmann}, {Harvanek}, {Krzesi{\'n}ski}, {Neilsen}, {Long}, {Schneider}, \&
  {Snedden}}]{Eis2006}
{Eisenstein}, D.~J., {Liebert}, J., {Harris}, H.~C., {et~al.} 2006, \apjs, 167,
  40

\bibitem[{{Fontaine} {et~al.}(2011){Fontaine}, {Brassard}, \&
  {Bergeron}}]{Fon2001}
{Fontaine}, G., {Brassard}, P., \& {Bergeron}, P. 2011, \pasp, 113, 409

\bibitem[{{Fukugita} {et~al.}(1996){Fukugita}, {Ichikawa}, {Gunn}, {Doi},
  {Shimasaku}, \& {Schneider}}]{Fukugita1996}
{Fukugita}, M., {Ichikawa}, T., {Gunn}, J.~E., {et~al.} 1996, \aj, 111, 1748

\bibitem[{{Garc\'{i}a-Berro} {et~al.}(1999){Garc\'{i}a-Berro}, E., {Isern}, \&
  {Burkert}}]{Gar1999}
{Garc\'{i}a-Berro}, E., {Torres}, S., {Isern}, J., \& {Burkert}, A. 1999,
  \mnras, 302, 173

\bibitem[{{Garc\'{i}a-Berro} {et~al.}(1988){Garc\'{i}a-Berro}, {Hernanz},
  {Isern}, \& {Mochkovitch}}]{Gar1988}
{Garc\'{i}a-Berro}, E., {Hernanz}, M., {Isern}, J., \& {Mochkovitch}, R. 1988,
  \nat, 333, 642

\bibitem[{{Garcia-Berro} {et~al.}(1995){Garcia-Berro}, {Hernanz}, {Isern}, \&
  {Mochkovitch}}]{gdot}
{Garcia-Berro}, E., {Hernanz}, M., {Isern}, J., \& {Mochkovitch}, R. 1995,
  \mnras, 277, 801

\bibitem[{{Garc{\'{\i}}a-Berro} {et~al.}(2011){Garc{\'{\i}}a-Berro},
  {Lor{\'e}n-Aguilar}, {Torres}, {Althaus}, \& {Isern}}]{JCAP}
{Garc{\'{\i}}a-Berro}, E., {Lor{\'e}n-Aguilar}, P., {Torres}, S., {Althaus},
  L.~G., \& {Isern}, J. 2011, \jcap, 5, 21

\bibitem[{{Garc\'{i}a-Berro} {et~al.}(2010){Garc\'{i}a-Berro}, {Torres},
  {Althaus}, {Renedo}, {Lor\'{e}n-Aguilar}, {C\'{o}rsico}, {Rohrmann},
  {Salaris}, \& {Isern}}]{Gar2010}
{Garc\'{i}a-Berro}, E., {Torres}, S., {Althaus}, L.~G., {et~al.} 2010, \nat,
  465, 194

\bibitem[{{Garc\'{i}a-Berro} {et~al.}(2004){Garc\'{i}a-Berro}, {Torres},
  {Isern}, \& {Burkert}}]{Gar2004}
{Garc\'{i}a-Berro}, E., {Torres}, S., {Isern}, J., \& {Burkert}, A. 2004, \aa,
  418, 53

\bibitem[{{Geijo} {et~al.}(2006){Geijo}, {Torres}, {Isern}, \&
  {Garc{\'{\i}}a-Berro}}]{Geijo1}
{Geijo}, E.~M., {Torres}, S., {Isern}, J., \& {Garc{\'{\i}}a-Berro}, E. 2006,
  \mnras, 369, 1654

\bibitem[{{Giammichele} {et~al.}(2012){Giammichele}, {Bergeron}, \&
  {Dufour}}]{local}
{Giammichele}, N., {Bergeron}, P., \& {Dufour}, P. 2012, \apjs, 199, 29

\bibitem[{{Gunn} {et~al.}(1998{\natexlab{a}}){Gunn}, {Carr}, {Rockosi},
  {Sekiguchi}, {Berry}, {Elms}, {de Haas}, {Ivezi{\'c}}, {Knapp}, {Lupton},
  {Pauls}, {Simcoe}, {Hirsch}, {Sanford}, {Wang}, {York}, {Harris}, {Annis},
  {Bartozek}, {Boroski}, {Bakken}, {Haldeman}, {Kent}, {Holm}, {Holmgren},
  {Petravick}, {Prosapio}, {Rechenmacher}, {Doi}, {Fukugita}, {Shimasaku},
  {Okada}, {Hull}, {Siegmund}, {Mannery}, {Blouke}, {Heidtman}, {Schneider},
  {Lucinio}, \& {Brinkman}}]{Gun1998}
{Gunn}, J.~E., {Carr}, M., {Rockosi}, C., {et~al.} 1998{\natexlab{a}}, \aj,
  116, 3040

\bibitem[{{Gunn} {et~al.}(1998{\natexlab{b}}){Gunn}, {Carr}, {Rockosi},
  {Sekiguchi}, {Berry}, {Elms}, {de Haas}, {Ivezi{\'c}}, {Knapp}, {Lupton},
  {Pauls}, {Simcoe}, {Hirsch}, {Sanford}, {Wang}, {York}, {Harris}, {Annis},
  {Bartozek}, {Boroski}, {Bakken}, {Haldeman}, {Kent}, {Holm}, {Holmgren},
  {Petravick}, {Prosapio}, {Rechenmacher}, {Doi}, {Fukugita}, {Shimasaku},
  {Okada}, {Hull}, {Siegmund}, {Mannery}, {Blouke}, {Heidtman}, {Schneider},
  {Lucinio}, \& {Brinkman}}]{Gunn1998}
{Gunn}, J.~E., {Carr}, M., {Rockosi}, C., {et~al.} 1998{\natexlab{b}}, \aj,
  116, 3040

\bibitem[{{Gunn} {et~al.}(2006){Gunn}, {Siegmund}, {Mannery}, {Owen}, {Hull},
  {Leger}, {Carey}, {Knapp}, {York}, {Boroski}, {Kent}, {Lupton}, {Rockosi},
  {Evans}, {Waddell}, {Anderson}, {Annis}, {Barentine}, {Bartoszek}, {Bastian},
  {Bracker}, {Brewington}, {Briegel}, {Brinkmann}, {Brown}, {Carr},
  {Czarapata}, {Drennan}, {Dombeck}, {Federwitz}, {Gillespie}, {Gonzales},
  {Hansen}, {Harvanek}, {Hayes}, {Jordan}, {Kinney}, {Klaene}, {Kleinman},
  {Kron}, {Kresinski}, {Lee}, {Limmongkol}, {Lindenmeyer}, {Long}, {Loomis},
  {McGehee}, {Mantsch}, {Neilsen}, {Neswold}, {Newman}, {Nitta}, {Peoples},
  {Pier}, {Prieto}, {Prosapio}, {Rivetta}, {Schneider}, {Snedden}, \&
  {Wang}}]{Gunn2006}
{Gunn}, J.~E., {Siegmund}, W.~A., {Mannery}, E.~J., {et~al.} 2006, \aj, 131,
  2332

\bibitem[{{Hambly} {et~al.}(2001){Hambly}, {MacGillivray}, {Read}, {Tritton},
  {Thomson}, {Kelly}, {Morgan}, {Smith}, {Driver}, {Williamson}, {Parker},
  {Hawkins}, {Williams}, \& {Lawrence}}]{SSS}
{Hambly}, N.~C., {MacGillivray}, H.~T., {Read}, M.~A., {et~al.} 2001, \mnras,
  326, 1279

\bibitem[{{Hansen} {et~al.}(2002){Hansen}, {Brewer}, {Fahlman}, {Gibson},
  {Ibata}, {Limongi}, {Rich}, {Richer}, {Shara}, \& {Stetson}}]{Han2002}
{Hansen}, B.~M.~S., {Brewer}, J., {Fahlman}, G.~G., {et~al.} 2002, \apjl, 574,
  L155

\bibitem[{{Hansen} {et~al.}(2013){Hansen}, {Kalirai}, {Anderson}, {Dotter},
  {Richer}, {Rich}, {Shara}, {Fahlman}, {Hurley}, {King}, {Reitzel}, \&
  {Stetson}}]{HansenGC}
{Hansen}, B.~M.~S., {Kalirai}, J.~S., {Anderson}, J., {et~al.} 2013, \nat, 500,
  51

\bibitem[{{Harris} {et~al.}(2006){Harris}, {Munn}, {Kilic}, {Liebert},
  {Williams}, {von Hippel}, {Levine}, {Monet}, {Eisenstein}, {Kleinman},
  {Metcalfe}, {Nitta}, {Winget}, {Brinkmann}, {Fukugita}, {Knapp}, {Lupton},
  {Smith}, \& {Schneider}}]{Har2006}
{Harris}, H.~C., {Munn}, J.~A., {Kilic}, M., {et~al.} 2006, \aj, 131, 571

\bibitem[{{Haywood}(2008)}]{2008MNRAS.388.1175H}
{Haywood}, M. 2008, \mnras, 388, 1175

\bibitem[{{Isern} {et~al.}(2005){Isern}, {Garc{\'{\i}}a-Berro},
  {Dom{\'{\i}}guez}, {Salaris}, \& {Straniero}}]{2005ASPC..334...43I}
{Isern}, J., {Garc{\'{\i}}a-Berro}, E., {Dom{\'{\i}}guez}, I., {Salaris}, M.,
  \& {Straniero}, O. 2005, in Astronomical Society of the Pacific Conference
  Series, Vol. 334, 14th European Workshop on White Dwarfs, ed. D.~{Koester} \&
  S.~{Moehler}, 43

\bibitem[{{Isern} {et~al.}(2000){Isern}, {Garc{\'{\i}}a-Berro}, {Hernanz}, \&
  {Chabrier}}]{energy2}
{Isern}, J., {Garc{\'{\i}}a-Berro}, E., {Hernanz}, M., \& {Chabrier}, G. 2000,
  \apj, 528, 397

\bibitem[{{Isern} {et~al.}(1998){Isern}, {Garc\'{i}a-Berro}, {Hernanz},
  {Mochkovitch}, \& {Torres}}]{Ise1998}
{Isern}, J., {Garc\'{i}a-Berro}, E., {Hernanz}, M., {Mochkovitch}, R., \&
  {Torres}, S. 1998, \apj, 503, 239

\bibitem[{{Isern} {et~al.}(2008){Isern}, {Garc{\'{\i}}a-Berro}, {Torres}, \&
  {Catal{\'a}n}}]{Isern}
{Isern}, J., {Garc{\'{\i}}a-Berro}, E., {Torres}, S., \& {Catal{\'a}n}, S.
  2008, \apjl, 682, L109

\bibitem[{{Isern} {et~al.}(1991){Isern}, {Hernanz}, {Mochkovitch}, \&
  {Garcia-Berro}}]{neon1}
{Isern}, J., {Hernanz}, M., {Mochkovitch}, R., \& {Garcia-Berro}, E. 1991,
  \aap, 241, L29

\bibitem[{{Isern} {et~al.}(1997){Isern}, {Mochkovitch}, {Garcia-Berro}, \&
  {Hernanz}}]{energy1}
{Isern}, J., {Mochkovitch}, R., {Garcia-Berro}, E., \& {Hernanz}, M. 1997,
  \apj, 485, 308

\bibitem[{James(1990)}]{James_1990}
James, F. 1990, Comput. Phys. Commun., 60, 329

\bibitem[{{Jeffery} {et~al.}(2011){Jeffery}, {von Hippel}, {DeGennaro}, {van
  Dyk}, {Stein}, \& {Jefferys}}]{Jeffery}
{Jeffery}, E.~J., {von Hippel}, T., {DeGennaro}, S., {et~al.} 2011, \apj, 730,
  35

\bibitem[{{Jordi} {et~al.}(2006){Jordi}, {Grebel}, \& {Ammon}}]{Jordi}
{Jordi}, K., {Grebel}, E.~K., \& {Ammon}, K. 2006, \aap, 460, 339

\bibitem[{{Kalirai} {et~al.}(2001){Kalirai}, {Ventura}, {Richer}, {Fahlman},
  {Durrell}, {D'Antona}, \& {Marconi}}]{Kal2001}
{Kalirai}, J.~S., {Ventura}, P., {Richer}, H.~B., {et~al.} 2001, \aj, 122, 3239

\bibitem[{{Kawaler}(1996)}]{Kawaler}
{Kawaler}, S.~D. 1996, \apjl, 467, L61

\bibitem[{{Kroupa}(2001)}]{Kro2001}
{Kroupa}, P. 2001, \mnras, 322, 231

\bibitem[{{Krzesinski} {et~al.}(2009){Krzesinski}, {Kleinman}, {Nitta},
  {H\"{u}gelmeyer}, {Dreizler}, {Liebert}, \& {Harris}}]{Krz2009}
{Krzesinski}, J., {Kleinman}, S.~J., {Nitta}, A., {et~al.} 2009, \aa, 508, 471

\bibitem[{{Mochkovitch} {et~al.}(1990){Mochkovitch}, {Garcia-Berro}, {Hernanz},
  {Isern}, \& {Panis}}]{1990A&A...233..456M}
{Mochkovitch}, R., {Garcia-Berro}, E., {Hernanz}, M., {Isern}, J., \& {Panis},
  J.~F. 1990, \aap, 233, 456

\bibitem[{{Noh} \& {Scalo}(1990)}]{NohScalo1990}
{Noh}, H.-R. \& {Scalo}, J. 1990, \apj, 352, 605

\bibitem[{{Nordstr{\"o}m} {et~al.}(2004){Nordstr{\"o}m}, {Mayor}, {Andersen},
  {Holmberg}, {Pont}, {J{\o}rgensen}, {Olsen}, {Udry}, \&
  {Mowlavi}}]{2004A&A...418..989N}
{Nordstr{\"o}m}, B., {Mayor}, M., {Andersen}, J., {et~al.} 2004, \aap, 418, 989

\bibitem[{{Olsen}(1983)}]{1983A&AS...54...55O}
{Olsen}, E.~H. 1983, \aaps, 54, 55

\bibitem[{{Pauli} {et~al.}(2003){Pauli}, {Napiwotzki}, {Altmann}, {Heber},
  {Odenkirchen}, \& {Kerber}}]{Pauli}
{Pauli}, E.-M., {Napiwotzki}, R., {Altmann}, M., {et~al.} 2003, \aap, 400, 877

\bibitem[{{Pietrinferni} {et~al.}(2004){Pietrinferni}, {Cassisi}, {Salaris}, \&
  {Castelli}}]{BaSTI}
{Pietrinferni}, A., {Cassisi}, S., {Salaris}, M., \& {Castelli}, F. 2004, \apj,
  612, 168

\bibitem[{{Press} {et~al.}(1986){Press}, {Flannery}, \& {Teukolsky}}]{NRs}
{Press}, W.~H., {Flannery}, B.~P., \& {Teukolsky}, S.~A. 1986, {Numerical
  Recipes. The art of scientific computing} (Cambridge: University Press, 1986)

\bibitem[{{Renedo} {et~al.}(2010){Renedo}, {Althaus}, {Miller Bertolami},
  {Romero}, {C{\'o}rsico}, {Rohrmann}, \& {Garc{\'{\i}}a-Berro}}]{Ren2010}
{Renedo}, I., {Althaus}, L.~G., {Miller Bertolami}, M.~M., {et~al.} 2010, \apj,
  717, 183

\bibitem[{{Rohrmann} {et~al.}(2012){Rohrmann}, {Althaus},
  {Garc{\'{\i}}a-Berro}, {C{\'o}rsico}, \& {Miller Bertolami}}]{Rene}
{Rohrmann}, R.~D., {Althaus}, L.~G., {Garc{\'{\i}}a-Berro}, E., {C{\'o}rsico},
  A.~H., \& {Miller Bertolami}, M.~M. 2012, \aap, 546, A119

\bibitem[{{Rowell}(2013)}]{Rowell}
{Rowell}, N. 2013, \mnras, 434, 1549

\bibitem[{{Rowell} \& {Hambly}(2011)}]{Row2011}
{Rowell}, N. \& {Hambly}, N.~C. 2011, \mnras, 417, 93

\bibitem[{{Salaris} {et~al.}(2013){Salaris}, {Althaus}, \&
  {Garc{\'{\i}}a-Berro}}]{Salaris13}
{Salaris}, M., {Althaus}, L.~G., \& {Garc{\'{\i}}a-Berro}, E. 2013, \aap, 555,
  A96

\bibitem[{Salaris {et~al.}(2010)Salaris, Cassisi, Pietrinferni, Kowalski, \&
  Isern}]{Sal2010}
Salaris, M., Cassisi, S., Pietrinferni, A., Kowalski, P.~M., \& Isern, J. 2010,
  \apj, 716, 1241

\bibitem[{{Schmidt}(1968)}]{Sch1968}
{Schmidt}, M. 1968, \apj, 151, 393

\bibitem[{{Segretain} {et~al.}(1994){Segretain}, {Chabrier}, {Hernanz},
  {Garcia-Berro}, {Isern}, \& {Mochkovitch}}]{segretain}
{Segretain}, L., {Chabrier}, G., {Hernanz}, M., {et~al.} 1994, \apj, 434, 641

\bibitem[{{Sellwood} \& {Binney}(2002)}]{Sel2002}
{Sellwood}, J.~A. \& {Binney}, J.~J. 2002, \mnras, 336, 785

\bibitem[{{Str\"omgren}(1987)}]{1987gal..proc..229S}
{Str\"omgren}, B. 1987, in NATO ASIC Proc. 207: The Galaxy, ed. G.~{Gilmore} \&
  B.~{Carswell}, 229--246

\bibitem[{{Torres} {et~al.}(2002){Torres}, {Garc\'{i}a-Berro}, {Burkert}, \&
  {Isern}}]{Tor2002}
{Torres}, S., {Garc\'{i}a-Berro}, E., {Burkert}, A., \& {Isern}, J. 2002,
  \mnras, 336, 971

\bibitem[{{Torres} {et~al.}(2007){Torres}, {Garc{\'{\i}}a-Berro}, \&
  {Isern}}]{Geijo2}
{Torres}, S., {Garc{\'{\i}}a-Berro}, E., \& {Isern}, J. 2007, \mnras, 378, 1461

\bibitem[{{Twarog}(1980)}]{Twarog1980}
{Twarog}, B.~A. 1980, \apj, 242, 242

\bibitem[{{Winget} {et~al.}(1987){Winget}, {Hansen}, {Liebert}, {van Horn},
  {Fontaine}, {Nather}, {Kepler}, \& {Lamb}}]{Win1987}
{Winget}, D.~E., {Hansen}, C.~J., {Liebert}, J., {et~al.} 1987, \apj, 315, L77

\bibitem[{{Xu} {et~al.}(2007){Xu}, {Deng}, \& {Hu}}]{xu2007}
{Xu}, Y., {Deng}, L.~C., \& {Hu}, J.~Y. 2007, \mnras, 379, 1373

\end{thebibliography}

\end{document}